 \documentclass[final,5p,times,twocolumn]{elsarticle}

\usepackage{amssymb}
\usepackage{amsmath}
\newtheorem{Definition}{Definition}
\usepackage[linesnumbered,noend, ruled]{algorithm2e}
\usepackage{subfigure}
\usepackage{booktabs}

\begin{document}

\begin{frontmatter}



\title{GAC: A Deep Reinforcement Learning Model Toward User Incentivization in Unknown Social Networks}


\author[label1,label2]{Shiqing Wu\corref{cor1}}
\author[label3]{Weihua Li}
\author[label2]{Quan Bai\corref{cor1}}

\affiliation[label1]{organization={School of Computer Science, University of Technology Sydney},
            country={Australia}}
\affiliation[label2]{organization={School of Information and Communication Technology, University of Tasmania},
            country={Australia}}
\affiliation[label3]{organization={School of Engineering, Computer and Mathematical Sciences, Auckland University of Technology},
            country={New Zealand}}
            
\cortext[cor1]{Corresponding Author. Email: shiqing.wu@uts.edu.au (S. Wu), weihua.li@aut.ac.nz (W. Li), quan.bai@utas.edu.au (Q. Bai). }

\begin{abstract}
In recent years, many applications have deployed incentive mechanisms to promote users' attention and engagement. Most incentive mechanisms determine specific incentive values based on users' attributes (e.g., preferences), while such information is unavailable in many real-world applications. Meanwhile, due to budget restrictions, realizing successful incentivization for all users can be challenging to complete. In this light, we consider leveraging social influence to maximize the incentivization result. We can directly incentivize influential users to affect more users, so the cost of incentivizing these users can be decreased. However, identifying influential users in a social network requires complete information about influence strength among users, which is impractical to acquire in real-world situations. In this research, we propose an end-to-end reinforcement learning-based framework, called Geometric Actor-Critic (GAC), to tackle the abovementioned problem. The proposed approach can realize effective incentive allocation without having prior knowledge about users' attributes. Three real-world social network datasets have been adopted in the experiments to evaluate the performance of GAC. The experimental results indicate that GAC can learn and apply effective incentive allocation policies in unknown social networks and outperform existing incentive allocation approaches. 
\end{abstract}



\end{frontmatter}


\section{Introduction}
In recent years, incentive mechanisms used to incentivize users to select behaviors beneficial to incentive providers have been widely studied and applied in many fields. Typical examples can be promoting sales \cite{Zhao2018Selling, Zhang2020Incentivize}, hiring workers \cite{Singer2013Pricing, Gan2017Incentivize, Qiu2019Incentivizing}, and encouraging beneficial behaviors \cite{Singla2015Incentivizing, Wu2018Greencommute}. Nevertheless, the effectiveness of incentives can vary between different scenarios, which mainly depends on the properties of both the scenarios and users, such as the purpose of incentivization and users' interests \cite{Truong2018Adaptive}. Hence, it is necessary to learn an effective incentive policy to maximize the effectiveness of incentives and incentivize more users under a limited budget. This process can be modeled as \emph{incentive allocation} problem \cite{Wu2021Learning}.

In general, an incentive provider tends to minimize the cost of providing an incentive such that the utility derived from the incentive can be maximized. In contrast, users may not accept the incentives offered since they also expect to maximize their profit. This conflict makes incentive allocation a challenging problem, as overpricing or underpricing the incentive may lead to overuse of budget or failure of incentivization \cite{Singla2013Truthful}. Therefore, an appropriate incentive structure and pricing policy for users can be considered a key factor for successful incentive allocation. More specifically, the pricing policy should not only maximize the incentive provider's utility but also maintain users' satisfaction. Some approaches attempt to provide ``optimal'' incentives to users by considering their attributes, such as personal preferences and skill abilities \cite{Gan2017Incentivize, Wu2018Greencommute, Wu2019C}. However, such information can be unavailable or incomplete in many real-world applications due to privacy concerns, making these attribute-based approaches ineffective. In this case, it is essential to propose an effective approach that does not rely on prior knowledge about users to allocate incentives.

Meanwhile, the limited budget may be challenging to provide sufficient incentives to all users, limiting the overall performance of user incentivization. Some studies suggest that incentivizing users can be facilitated by using social influence, as persuasive information can be quickly spread among friends in a social network \cite{Li2018Influence}. For example, many companies ``engage'' some influential users on social media by providing them with free products, with the expectation that these influential users can introduce and promote the products to their followers \cite{Lovett2013On}. By spreading influence in the social network, influential users can influence their followers and other users in the same network \cite{axsen2013Social}. In other words, providing appropriate incentives to some influential users can indirectly influence users' behaviors through social influence.

However, identifying influential users in a social network is challenging in real-world applications since the influence strength between each pair of users is difficult to estimate. Although a user's influential capabilities can be estimated simply by the number of followers they have, users with many fake followers usually have very limited influence \cite{Wilder2018Maximizing}. Moreover, the strength with which influential users influence their friends or followers may vary with respect to different items or topics \cite{Tang2009social}. For example, a famous blogger who focuses on commenting on movies may have a stronger influence on her fans' movie choices but little influence on music choices. In principle, comprehensive surveys can be conducted on each potentially influential user to estimate the strength of influence between each pair of users. However, it is impractical to investigate all users in a social network because of its huge cost \cite{Valente2007Identifying}.

Considering the abovementioned challenges, how can we learn effective strategies for allocating incentives in a social network with a limited budget, where there is no knowledge about users and the strength of influence among users? This scenario is widespread in real-world applications. When an incentive provider intends to incentivize users' behaviors on social media, it is difficult to obtain complete information about users and the network due to privacy and protocol restrictions. Therefore, a novel approach is required to learn how to allocate incentives effectively in such scenarios. To address this problem, we consider using reinforcement learning (RL) to explore policies for allocating incentives in a given network topology. More specifically, the RL agent can automatically adjust and determine the policy based on the interaction with users. To better utilize the budget, the RL agent also learns the network representation to realize effective incentive allocation by recognizing influential users in the network to incentivize them strategically.

In our previous work \cite{Wu2021Learning}, we conducted preliminary research on using reinforcement learning to address the incentive allocation problem in social networks with limited information. However, the details of applying reinforcement learning to the incentive allocation problem have not been discussed. In this paper, we first formulate the incentive allocation problem as a sequential decision problem. Then, we propose a novel end-to-end reinforcement learning-based approach called Geometric Actor-Critic (GAC), which can extract information from user behaviors and the social network to learn representations for each user and the network, respectively. Benefiting from this, the RL agent can identify the potential influential users in the network, and then generate incentive allocation policies. In the experiments, three real-world social network datasets are used to build the simulation environment for the evaluation. The experimental results show that the proposed GAC outperforms existing approaches for incentive allocation in incentivizing users in unknown social networks. 

The major contributions of this research work can be summarized as follow:
\begin{itemize}
\item We adopt Markov Decision Process (MDP) to model the incentive allocation problem as a sequential decision problem, which makes using reinforcement learning to tackle the incentive allocation problem possible.

\item We propose a novel reinforcement learning framework to learn effective incentive allocation policies in unknown social networks, i.e., Geometric Actor-Critic (GAC). A key feature of GAC is that it only requires basic information about the social network (i.e., the topology of the network) and the observation of users' behaviors for training and exploitation. Unlike most existing approaches for incentive allocation, the proposed GAC can learn effective policies for incentive allocation without knowing users' attributes beforehand (e.g., users' preferences). To the best of our knowledge, this is the first work that employs reinforcement learning to tackle the incentive allocation problem in unknown social networks.

\item We deploy three real-world social network datasets to evaluate the proposed GAC. The experimental results demonstrate that GAC outperforms the compared approaches and prove its effectiveness.
\end{itemize}

The remainder of this paper is organized as follows: Section \ref{sec:relatedwork} reviews literature related to this research work. In Section \ref{sec:preliminaries}, we first introduce the problem of incentive allocation in unknown networks. Then we present notations and formal definitions used in this paper. Meanwhile, we introduce the simulation environment for training and evaluating the RL-based models. Subsequently, we give details about the proposed reinforcement learning-based framework in Section \ref{sec:methodology}. Sections \ref{sec:expsetup} and \ref{sec:expresult} demonstrate the experimental setup and results, respectively. Section \ref{sec:conclusion} illustrates the conclusion and future direction.

\section{Related Work}\label{sec:relatedwork}
Effective approaches for incentive allocation to motivate user behavior have been widely studied in recent years. These studies aim to address the problem of maximizing user incentivization under a budget constraint. Some incentive allocation approaches consider utilizing users' features and attributes to determine incentive values, such as users' preferences, location, and skill abilities \cite{Gan2017Incentivize, Sengvong2017Persuasive, Qiu2019Incentivizing, Xiao2019Model, Li2019Redundancy, Wu2021Identifying}. Although these approaches can effectively allocate incentives, their performance cannot be sustained if the information they rely on is unavailable. Meanwhile, counterfactual inference-based approaches that generate incentives by learning logged feedback data from users are also widely studied \cite{ Wu2019Incentivizeing, Lopez2020Cost}. A typical example is the budgeted multi-armed bandit problem, in which price options are modeled as arms, and the goal is to select the most advantageous arm \cite{Tran-Thanh2010Epsilon, Tran-Thanh2012Knapsack, Xia2015Thompson}. To apply such approaches in the real world, Singla et al. designed a UCB-based pricing mechanism called DBP-UCB to incentivize bike-sharing users to help reallocate bikes to suitable stations \cite{Singla2015Incentivizing}. However, these approaches assume that user behavior is independent and ignore user interaction and the influence between each other.

As aforementioned, a limited budget would restrict the number of users who can receive sufficient incentives and then be incentivized. To incentivize more users with a given budget, many studies have considered utilizing social influence that naturally exists among users to indirectly incentivize users \cite{Wu2018Greencommute, Zhao2018Selling, Zhang2020Incentivize}. These works assumed that the information about the social network, including the network topology and the influence strength of each pair of users, is known. Based on this assumption, these approaches can accurately locate influential users and incentivize them in priority. However, in the practical application, it is difficult to harvest complete information about the social network \cite{Wilder2018Maximizing}, which makes identifying influential users challenging. Several studies have focused on a similar problem called influence maximization in unknown networks, where the network structural information is incomplete or lacking. The objective is to identify a set of users who can maximize influence diffusion \cite{ Lei2015Online, Mihara2015Influence, Eshghi2019Efficient}. The difference between our problem and theirs is that they do not consider the cost of incentivizing each candidate user while we need to do. 

In recent years, more and more studies have started utilizing deep reinforcement learning to tackle some interesting challenges in real-world applications. For example, Xiao et al. proposed a DQN-based approach to realize optimal payment policy in a mobile crowdsourcing environment \cite{Xiao2017Secure}. Huang et al. utilized deep reinforcement learning to enhance the performance of recommender systems on the long-term recommendation accuracy \cite{Huang2021Deep}. Jahanshahi et al. adopted deep reinforcement learning to solve the meal delivery problem \cite{Jahanshahi2022Deep}. In addition, some studies have used deep reinforcement learning to solve combinatorial problems on graphs. Liu et al. proposed a dynamic knowledge graph reasoning framework based on deep reinforcement learning \cite{Liu2022Dynamic}. Yu et al. proposed a deep reinforcement learning-based approach to generate an optimal strategy for determining vehicle routing plans with minimal computation time \cite{Yu2019Online}. Bello et al. focused on tackling the traveling salesman problem by using reinforcement learning models \cite{Bello2017Neural}. Kamarthi et al. utilized DQN to realize efficient graph sampling to tackle the influence maximization problem \cite{Kamarthi2020Influence}. In contrast to these works, our model aims to solve the incentive allocation problem in a scenario where only the social network topology is given. Our goal is to train an effective incentive allocation policy that can generate incentives for each user subject to a budget constraint. To the best of our knowledge, this is the first work that employs deep reinforcement learning to tackle the incentive allocation problem in unknown networks.

\section{Preliminaries}\label{sec:preliminaries}

\begin{figure}[t]
  \centering
  \includegraphics[width=0.8\linewidth]{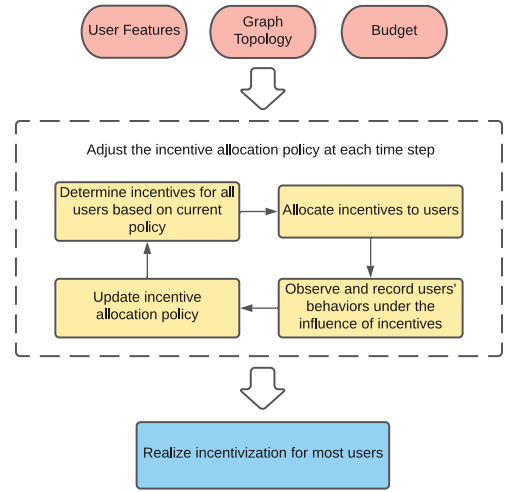}
  \caption{Learning policies for incentive allocation}
  \label{fig:problem}
\end{figure}

\subsection{Problem Description}
User incentivization is a complex problem, as the user's decision-making can be affected by external factors, such as incentives. To effectively incentivize users' behaviors, applying a proper incentive allocation policy is essential. In this paper, the problem we focus on is to incentivize as many users as possible in a social network under a limited budget restriction. The information about the network is the topology only, and no prior knowledge about users is given. To tackle this problem, it is necessary to learn the right policy for incentive allocation, which simultaneously considers users' attitudes toward incentives and their influential roles in the social network. The goal is that the approach eventually can incentivize users with appropriate incentives and tends to incentivize influential users in priority. 

Figure \ref{fig:problem} shows a general process of incentive allocation. Given the network's topology and a fixed budget per time step, the system allocates incentives to all users based on the current allocation policy at every time step. At each time step, the approach would update its parameters based on observing users' behaviors, i.e., whether the user accepts the incentive or not. This whole process repeats for a finite number of steps. After the last time step, the number of incentivized users reflects the performance of the approach. A typical example of this process is that a business company advertises its product on social media, expecting many users purchase it. The business company would tend to hire some influential users to propagate information about its product and then provide each non-influential user a minor incentive, such as coupons or discounts. To make use of the limited budget efficiently, the incentive allocation process should be able to produce maximum profits, i.e., successfully incentivize more users. However, since the company cannot obtain complete information about users' profiles and the entire social network, it is necessary to incentivize users using different incentives to identify the optimal incentive allocation policy. Therefore, proposing an effective and adaptive approach to adjust incentive allocation policies dynamically and autonomously is crucial to the incentive allocation problem.

\subsection{Formal Definitions}
In this section, we introduce definitions and notations used in this paper. Notations are listed in Table \ref{tab:notations}.

\begin{table}[!t]
\centering
\caption{Notations used in this paper}
\label{tab:notations}
\resizebox{\columnwidth}{!}{
\begin{tabular}{ll}
\hline
Notation               & Description                                                       \\ \hline
$v_i$                  & A user $v_i$ in a social network                                  \\
$b_{v_i,t}$            & $v_i$'s behavior at time step $t$                                 \\
$G$                    & A directed graph of a social network                              \\
$V$                    & A set of users in $G$                                             \\
$e_{ij}$               & Influential relationship between $v_i$ and $v_j$                  \\
$E$                    & A set of edges in $G$                                             \\
$w_{ij}$               & Influence strength associated on $e_{ij}$                         \\
$N_{v_i}^{in}$         & A set of $v_i$'s one-hop in-neighbors                             \\
$N_{v_i}^{out}$        & A set of $v_i$'s one-hop out-neighbors                            \\
$z_m$                  & A behavior option                                                 \\
$z^*$                  & The behavior that the incentive provider expects                  \\
$Z$                    & A set of all behavior options                                     \\
$p_{v_i,z_m}$          & $v_i$'s preference toward $z_m$                                   \\
$o_{v_i,t}$            & An incentive provided to $v_i$ at time step $t$                   \\
$B_t$                  & Remaining budget at time step $t$                                 \\
$\boldsymbol{A^{in}}$  & The in-adjacency matrix                                           \\
$\boldsymbol{A^{out}}$ & The out-adjacency matrix                                          \\
$\boldsymbol{F}$       & The feature matrix                                                \\
$\boldsymbol{f_{v_i}}$ & The feature vector for $v_i$                                                 \\
$u_t(v_i,z_m)$         & $v_i$'s utility toward $z_m$ at time step $t$                     \\
$k_t(v_i,z_m)$         & Influence from $v_i$'s in-neighbors toward $z_m$ at time step $t$ \\
$S_t$                  & State at time step $t$                                            \\
$\boldsymbol{a_t}$     & The action for the incentive allocation at time step $t$  \\
$R_t$                  & Step reward at time step $t$                                      \\
$r_{v_i,t}$            & Intermediate reward generated by $v_i$                            \\ \hline
\end{tabular}
}
\end{table}

\begin{Definition}\label{def:user}
An autonomous user agent representing a human user in a social network is defined as $v_i$. $b_{v_i,t}$ denotes $v_i$'s behavior at time step $t$.
\end{Definition}

\begin{Definition}\label{def:network}
A social network is represented as a directed graph $G=(V,E)$, where $V=\{v_1,...,v_i\}$ denotes a set of users, and $E=\{e_{ij}|\{v_i,v_j\}\subseteq V, i\neq j\}$ denotes a set of edges in the network. In $G$, $e_{ij}$ represents an influential relationship between $v_i$ and $v_j$. It implies that $v_i$ can influence $v_j$'s behavior, and the influence strength associated on $e_{ij}$ is represented as $w_{ij}\in(0,1]$. We use $N^{in}_{v_i}$ to represent a set of $v_i$'s one-hop in-neighbors who can influence $v_i$ directly. Similarly, $N^{out}_{v_i}$ represents a set of $v_i$'s one-hop out-neighbors who are directly influenced by $v_i$.
\end{Definition}

\begin{Definition}\label{def:item}
A behavior option $z_m \in Z$ represents a specific behavior that $v_i$ could choose, i.e., $b_{v_i, t}\in Z$, where $Z=\{z_1,z_2,...,z_m\}$ denotes a finite option set. The finite set of options $z_m$ can be a set of commercial products for users to purchase (e.g., coke or sprite) or actions that users can take (e.g., riding a bike or walking). For each $z_m$, $v_i$ has a personal preference $p_{v_i, z_m} \in [0,1]$. Higher $p_{v_i, z_m}$ implies that $v_i$ prefers to choose $z_m$, whereas smaller $p_{v_i, z_m}$ implies conversely. To clarify, we set the behavior that the incentive provider expects $v_i$ to select as $z^* \in Z$.
\end{Definition}

\begin{Definition}\label{def:incentive}
An incentive $o_{v_i,t}$ denotes the incentive provided to user $v_i$ from an incentive provider at time step $t$. The range of $o_{v_i,t}$ can be $[0,1]$. Note that the value of $o_{v_i,t}$ is strictly constrained by the remaining budget $B_t$, i.e., $o_{v_i,t} \leq B_t$.
\end{Definition}

In addition, several matrices will be used in section \ref{sec:methodology}. By default, we use bold upper and lowercase letters, e.g., $\boldsymbol{A}$ and $\boldsymbol{x}$ and to represent matrices and vectors. Let $\boldsymbol{A}^{out} \in \mathbb{R}^{|V|\times |V|}$ be out-adjacency matrix, where each entry $\boldsymbol{A}^{out}[i,j]=1$ if $e_{ij} \in E$. Similarly, we use $\boldsymbol{A}^{in}\in \mathbb{R}^{|V|\times |V|}$ to represent in-adjacency matrix, where each entry $\boldsymbol{A}^{in}[j,i]=1$ if $e_{ji} \in E$. The features matrix of all users is represented by $\boldsymbol{F}$, where each row represents a specific user's feature $\boldsymbol{f}_{v_i}=[o_{v_i,t}, b_{v_i,t}]$. In this paper, we use the one-hot representation for $b_{v_i,t}$.

\subsection{Simulation of Environment}
To model the process of user incentivization, we build a simulation-based environment by adopting the Agent-based Decision-making Model (ADM) \cite{Wu2019Adaptive}, in which two types of agents are involved, i.e., the RL agent and user agents. The RL agent stands for the incentive provider and generates incentives for users. In contrast, user agents represent human users in the real world. All user agents are required to behave at every time step, and each user agent always chooses an item with the highest user utility. Equation \ref{equ:behavior} formulates this behavior rule, where $u_t(v_i,z_m)$ denotes the user utility of $z_m$.

\begin{equation}\label{equ:behavior}
b_{v_i,t} = \arg\max_{z_m\in Z}u_t(v_i,z_m)
\end{equation}

The user utility $u_t(v_i,z_m)$, representing $v_i$'s satisfaction to $z_m$, is formulated in Equation \ref{equ:user_utility}, which consists of the user's preferences $p_{v_i, z_m}$, received incentive $o_{v_i,t}$, and the social influence $k_t(v_i, z_m)$ from its neighbors simultaneously. Higher user utility implies the option $z_m$ is more likely to be selected by $v_i$. Incentive $o_{v_i,t}$ would affect the user utility of $z^*$ only. Namely, users select items purely based on their preferences and social influence if no incentive is provided. This utility function makes users' behaviors in the simulation environment diverse since users may have different preferences and receive different social influences.

\begin{equation}\label{equ:user_utility}
u_t(v_i, z_m)=
\left\{
	\begin{array}{lr}
		p_{v_i, z_m}+k_t(v_i,z_m)+o_{v_i,t}, & z_m = z^*\\
		p_{v_i, z_m}+k_t(v_i,z_m), & otherwise
	\end{array}
\right.
\end{equation}

In the ADM, $b_{v_i,t}$ can produce influence on $b_{v_j,t+1},\forall v_j \in N^{out}_{v_i}$. Hence, multiple influences for supporting different items may co-exist simultaneously, which could form the conflict in affecting the focal user's behavior. Equation \ref{equ:influence} describes how the ADM formulates the social influence exerted on $v_i$ at time step $t$, where $w_{ji}$ denotes the strength of influence associated with edge $e_{ji}$.
\begin{equation}\label{equ:influence}
k_t(v_i, z_m)=\sum_{\substack{v_j\in N^{in}_{v_i},\\ b_{v_j,t-1}=z_m }}{w_{ji}}
\end{equation}

Although the ADM requires knowledge about users' preferences and the strength of influence associated with all edges to conduct the simulation, the proposed algorithm does not require such prior knowledge.

\section{Methodology}\label{sec:methodology}
In this section, we present the proposed approach for training an agent that can automatically generate effective incentive allocation policies. First, we formulate the incentive allocation problem as a Markov Decision Process (MDP) with a carefully designed reward function so that the training process can be adapted to different networks and budget constraints. We then present the structure of the proposed GAC and explain how we encode local and global embeddings as a state that provides contextual information to the RL agent to determine the incentive allocation policy. Finally, we introduce the process for training the proposed approach.

\subsection{Markov Decision Process Formulation for Incentive Allocation Problem}
First, we formulate the incentive allocation problem as an MDP. As aforementioned, the incentive allocation process goes through a series of time steps before it is completed. At each time step, the RL agent generates incentives for all users based on the current state of the network. After the incentives are allocated to all users, each user would decide whether to accept the incentive or not and provide feedback to the RL agent. The state of the network then changes at the end of the time step. Therefore, the incentive allocation problem can be modeled as a sequential decision problem. In this MDP, the corresponding formal representation of the transition at each time step can be formulated as $(S_t,\boldsymbol{a}_t,S_{t+1},R_t)$, where $S_t$ denotes the current state of the network, $\boldsymbol{a}_t$ represents the action, i.e., incentives allocated to all users, $S_{t+1}$ is the next state after incentives are provided, and $R_t$ is the step reward.

Notably, the action $\boldsymbol{a}_t$ is a vector where each element is a continuous value from 0 to 1 that needs to be converted to an incentive allocated to a specific user. The step reward $R_t$ is a cumulative reward formulated in Equation \ref{equ:stepreward}, where $r_{v_i,t}$ denotes the intermediate reward for assigning an incentive to $v_i$. The step reward is only available at the end of each time step.

\begin{equation}\label{equ:stepreward}
	R_t=\sum_{v_i\in V}{r_{v_i,t}}
\end{equation}

\begin{figure}[t]
  \centering
  \includegraphics[scale=0.6]{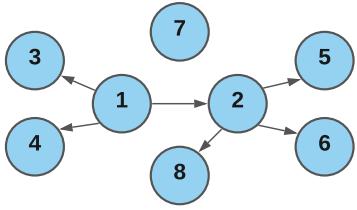}
  \caption{An example of a simple social network}
  \label{fig:smallnet}
\end{figure}

An intermediate reward $r_{v_i,t}$ is calculated by using Equation \ref{equ:interreward}. In the equation, $\alpha$ is the indicator of user activation. $\alpha=1$ if user $v_i$ accepts the incentive and choose $z^*$, and $\alpha=-1$ if conversely. This equation consists of two parts. The former part indicates the importance of $v_i$ in a social network, where $|N^{out}_{v_i}|$ and $|N^{in}_{v_i}|$ represent the number of neighbors who can be influenced by $v_i$ and who can influence $v_i$, respectively. To better explain the former part, we take the social network in Figure \ref{fig:smallnet} as an example. In this network, each node represents a specific user, and each directed edge represents the influence direction. Empirically, if a user can influence more users in the network, we can regard this user as more valuable to be incentivized. However, although users 1 and 2 can influence three users, incentivizing user 1 first could be a better choice when the budget is insufficient for incentivizing both of them, as user 1 can influence user 2. Based on this rule, the importance of each user should be calculated and distinguished. The latter part represents the efficiency of spending the budget. If a user can be incentivized with fewer incentives, then the budget can be saved for incentivizing subsequent users. Different from the step reward, the intermediate reward would be generated once $v_i$ responds to the provided incentive.

\begin{equation}\label{equ:interreward}
	r_{v_i,t}=\alpha(1 + \frac{|N^{out}_{v_i}|-|N^{in}_{v_i}|}{|V|})+\frac{\alpha+1}{2}\cdot \frac{B-o_{v_i,t}}{B}
\end{equation}

Algorithm \ref{alg:1} describes how the RL agent incentivizes users at every time step. The input is an action $\boldsymbol{a}_t$ generated by the RL agent and a fixed budget $B$, and the output is step reward $R_t$ and the log $L_t$, which is used to store records of users' behaviors. Line 1 sets the remaining budget as $B$ and the step reward as 0, and Line 2 initializes $L_t$ as an empty list. Lines 4-6 first map the value of incentive for $v_i$ from $\boldsymbol{a}_t$, and then check if the remaining budget $B_t$ is sufficient for allocating the incentive. In Lines 7-8, $v_i$ makes a decision and takes the behavior based on its user utilities. Subsequently, variables are updated based on the user's behavior $b_{v_i,t}$ in Lines 9-13. At last, in Lines 14-15, intermediate reward and step reward are calculated, and the record of $v_i$'s behavior is added to $L_t$ in Line 16.

\begin{algorithm}[t]
  \caption{Assign Incentives to users at time step $t$}
  \label{alg:1}
  \KwIn {Action $\boldsymbol{a}_t$, budget $B$}
  \KwOut {Step reward $R_t$, log $L_t$}
  	$B_t:=B$; $R_t:=0$;\\
  	Initialize $L_t$ as an empty list;\\
  	\For {$v_i \in V$}{
  		Map $o_{v_i,t}$ from $\boldsymbol{a}_t$;\\
  		\If{$o_{v_i,t}>B_t$}{
			$o_{v_i,t}=B_t$;\\  	
  		}
  		Calculate $u_t(v_i,z_m), \forall z_m \in Z$ by using Equation \ref{equ:user_utility};\\
  		$b_{v_i,t} = \arg\max\limits_{z_m\in Z}u_t(v_i,z_m)$;\\
  		\If{$b_{v_i,t} == z^*$}{
  			$B_t=B_t-o_{v_i,t}$;\\
  			$\alpha:=1$;\\
  		}
  		\Else{
			$\alpha:=-1$;\\  		
  		}
  		Calculate $r_{v_i,t}$ by using Equation \ref{equ:interreward};\\
  		$R_t=R_t+r_{v_i,t}$;\\
  		Add ($o_{v_i,t}, b_{v_i,t}$) to $L_t$;\\
  	}
	Return $R_t$, $L_t$
\end{algorithm} 

\subsection{Geometric Actor-Critic}
\begin{figure*}[!ht]
  \centering
  \includegraphics[width=\textwidth]{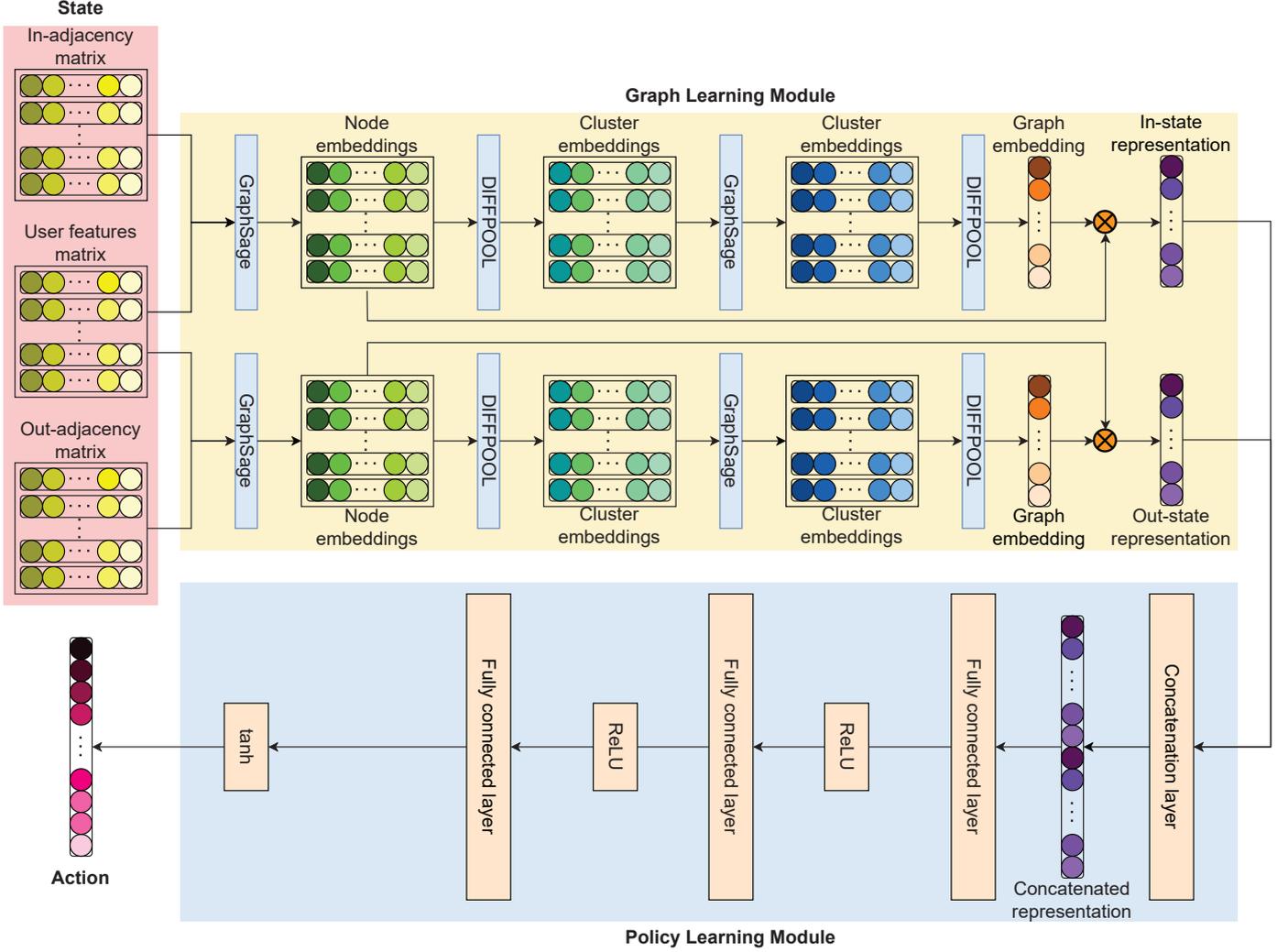}
  \caption{Architecture of GAC}
  \label{fig:structure}
\end{figure*}

To generate effective incentive allocation policies in social networks, the proposed approach should consider not only user features but also the network topology. Meanwhile, the proposed approach needs to encode the network from high-level to low-level representation due to the complexity and variable size of social networks.

In this case, we propose a reinforcement learning-based approach called Geometric Actor-Critic (GAC). Figure \ref{fig:structure} shows the structure of the proposed GAC. The GAC takes three matrices as input, i.e., the user feature matrix and two different adjacency matrices. These two adjacency matrices denote the in-adjacency matrix and the out-adjacency matrix. Since a social network is typically directed, its in-adjacency matrix is usually different from its out-adjacency matrix. These three matrices, i.e., user feature matrix, in-adjacency matrix, and out-adjacency matrix, form the state in the MDP.

The proposed GAC consists of two major modules, i.e., the graph learning module and the policy learning module. In the graph learning module, we first adopt two groups of Graph Neural Networks (GNNs) to encode the out-adjacency $\boldsymbol{A}^{out}$ and the in-adjacency matrices $\boldsymbol{A}^{in}$, respectively, to understand all information about the network. Each group of components starts from a GraphSage \cite{Hamilton2017Inductive}, aggregating user features based on the adjacency matrix to learn refined user features. In this work, we use the Mean ``variant'' of GraphSage by default. By taking the refining user features from the user's in-neighbors as an example, the formulation of a GraphSage can be formulated in Equations \ref{equ:graphsage1} and \ref{equ:graphsage2}. In this aggregation, $\boldsymbol{h}^q_{N^{in}_{v_i}}$ representing the vector representation of $N^{in}_{v_i}$ is first generated by aggregating representation of all $v_i$'s in-neighbors. Then, the GraphSage concatenates $v_i$'s representation $\boldsymbol{h}^{q-1}_{v_i}$ and $\boldsymbol{h}^q_{N^{in}_{v_i}}$. Subsequently, the concatenated vector is fed through a fully connected layer with a weight matrix $\boldsymbol{W}^q$ and a nonlinear activation function $ReLU$ to transform the representation of the user $v_i$. In this equation, $q$ denotes $q$-hop neighbors considered in the GraphSage. For example, when $q=1$, it means we only consider aggregating features from $v_i$'s one-hop neighbors, and then $\boldsymbol{h}^{q-1}_{v_i}$ becomes $\boldsymbol{h}^{0}_{v_i}$ representing the input user's feature. Note that the output from the first GraphSage is a matrix representing node embeddings for all users. This matrix would be fed to the subsequent DIFFPOOL.

\begin{equation}\label{equ:graphsage1}
\boldsymbol{h}^q_{N^{in}_{v_i}} \leftarrow MEAN_q(\{\boldsymbol{h}^{q-1}_{v_j}, v_j \in N^{in}_{v_i}\})
\end{equation}
\begin{equation}\label{equ:graphsage2}
\boldsymbol{h}^q_{v_i} \leftarrow ReLU(\boldsymbol{W}^q \cdot CONCAT(\boldsymbol{h}^{q-1}_{v_i}, \boldsymbol{h}^q_{N^{in}_{v_i}}))
\end{equation}

In the GAC, DIFFPOOL is used to learn the representation of the input graph by aggregating user features \cite{Ying2018Hierarchical}. DIFFPOOL can learn hierarchical representations of the input network by iteratively coarsening the network and mapping users to a set of clusters. DIFFPOOL first generates the embedding matrix $\boldsymbol{Y}^l \in \mathbb{R}^{n_l \times d}$ and the assignment matrix $\boldsymbol{V}^l \in \mathbb{R}^{n_l \times n_{l+1}}$, where $l$ denotes the index of DIFFPOOL, $n_l$ denotes the number of clusters or users in the input coarsened network, $n_{n+1}$ denotes the number of clusters in the output coarsened network, and $d$ denotes the dimension of embeddings. Equations \ref{equ:diffpool_emb} and \ref{equ:diffpool_pool} formulate how to generate $\boldsymbol{Y}^l$ and $\boldsymbol{V}^l$ via two independent GNNs, where $\boldsymbol{A}^l$ and $\boldsymbol{X}^l$ represent the corresponding adjacency matrix and node embeddings matrix. Here, we use two GraphSages for GNNs. Note that when $l=0$, $\boldsymbol{A}^l$ is the original adjacency matrix, and $\boldsymbol{X}^l$ denotes the node embeddings generated by the first GraphSage.

\begin{equation}\label{equ:diffpool_emb}
\boldsymbol{Y}^{l} = GNN_{embed}^{l}(\boldsymbol{A}^l,\boldsymbol{X}^l)
\end{equation}
\begin{equation}\label{equ:diffpool_pool}
\boldsymbol{V}^{l} = GNN_{pool}^{l}(\boldsymbol{A}^l,\boldsymbol{X}^l)
\end{equation}

Then, we can obtain a new embeddings matrix $\boldsymbol{X}^{l+1} \in \mathbb{R}^{n_{l+1} \times d}$ for each cluster in the coarsened network and a new coarsened adjacency matrix $\boldsymbol{A}^{l+1} \in \mathbb{R}^{n_{l+1} \times n_{l+1}}$ by using Equations \ref{equ:diffpool_X} and \ref{equ:diffpool_A}. The output $\boldsymbol{X}^{l+1}$ and $\boldsymbol{A}^{l+1}$ would be fed to the next GraphSage and then a DIFFPOOL.

\begin{equation}\label{equ:diffpool_X}
\boldsymbol{X}^{l+1} = V^{l^\top} Y^{l}
\end{equation}
\begin{equation}\label{equ:diffpool_A}
\boldsymbol{A}^{l+1} = V^{l^\top} Y^{l} V^l
\end{equation}

At last, the coarsened network would only have one cluster, i.e., a vector representation for the entire network. In this study, we use the first DIFFPOOL to map all users into 16 clusters and the second DIFFPOOL to generate the global representation for the corresponding network.
 
After obtaining the local and global representations, i.e., node embeddings for all users and graph embedding, we combine these two representations by using a matrix-vector product as described in Equations \ref{equ:comb_in} and \ref{equ:comb_out}. In these two equations, $\boldsymbol{x}_{in} \in \mathbb{R}^{1 \times d}$ and $\boldsymbol{x}_{out} \in \mathbb{R}^{1 \times d}$ represent graph embeddings generated by the in-adjacency and the out-adjacency matrices, respectively. While $\boldsymbol{H}_{in} \in \mathbb{R}^{|V| \times d}$ and $\boldsymbol{H}_{out} \in \mathbb{R}^{|V| \times d}$ represent the corresponding node embeddings. $d$ denotes the dimensional length of embedding. The matrices of node embeddings need to be transposed to make the calculation feasible. Through Equations \ref{equ:comb_in} and \ref{equ:comb_out}, two new vectors $\boldsymbol{x}^*_{in} \in \mathbb{R}^{1 \times |V|}$ and $\boldsymbol{x}^*_{out} \in \mathbb{R}^{1 \times |V|}$ which encode local and global representations together can be generated. 

\begin{equation}\label{equ:comb_in}
\begin{aligned}
\boldsymbol{x}^*_{in} = \boldsymbol{x}_{in}\boldsymbol{H}_{in}^{\top}
\end{aligned}
\end{equation}
\begin{equation}\label{equ:comb_out}
\boldsymbol{x}^*_{out} = \boldsymbol{x}_{out} \boldsymbol{H}_{out}^{\top}
\end{equation}

Subsequently, $\boldsymbol{x}^{*}_{in}$ and $\boldsymbol{x}^{*}_{out}$ would be normalized by using their L2-norm~\cite{Chopra2007Dynamics}. These two normalized vectors would be fed to a concatenation layer in the policy learning module. Then the concatenated vector is passed throughthree fully connected layers. The first two fully connected layers are equipped with $ReLU$ as the activation function, while the last fully connected layer uses $tanh$ instead. After rescaling the output action representation $\boldsymbol{a}_t \in \mathbb{R}^{1\times |V|}$ to the range from 0 to 1, the system can allocate incentives to all users if the budget is sufficient, and the entire GAC can be deployed as an end-to-end system. The generated action $\boldsymbol{a}_t$ would be converted to incentives for allocation to all users. Users' behaviors at time step $t$ would form the reward $R_t$ and the next state $S_{t+1}$. Finally, the transition $(S_t, \boldsymbol{a}_t,S_{t+1},R_t)$ would be store in the replay buffer.

The proposed GAC's computational complexity is determined by the two major modules, i.e., the graph learning module and the policy learning module. In the graph learning module, the inputted matrices would be fed into two sequences consisting of two GraphSages and two DIFFPOOLs. The computational complexity of the first GraphSage and the first DIFFPOOL are $\mathcal{O}(d^2\sum_{v_i\in V}|N_{v_i}^{in}|)$ (it would turn to $\mathcal{O}(d^2\sum_{v_i\in V}|N^{out}_{v_i}|)$ if out-adjacency matrix is processed) and $\mathcal{O}(|V|^2)$, where $d$ denotes the dimensions of user features, $|V|$ is the number of users in the network, and $|N_{v_i}|^{in}$ and $|N_{v_i}|^{out}$ represent the number of in-neighbors and out-neighbors of $v_i$. Regarding the second GraphSage and DIFFPOOL, their computational complexity would significantly decrease, as user embeddings would be aggregated into cluster embeddings. The complexity of the matrix-vector product is $\mathcal{O}(d|V|)$. In the policy learning module, the computational complexity of a fully connected layer is $\mathcal{O}(d^2)$. Hence, the overall computational complexity of the proposed GAC is $\mathcal{O}(|V|^2)$, as $\sum_{v_i\in V}|N_{v_i}^{in}| \le |V|^2$.

\subsection{Model Training}
\begin{figure*}[!t]
  \centering
  \includegraphics[width=\textwidth]{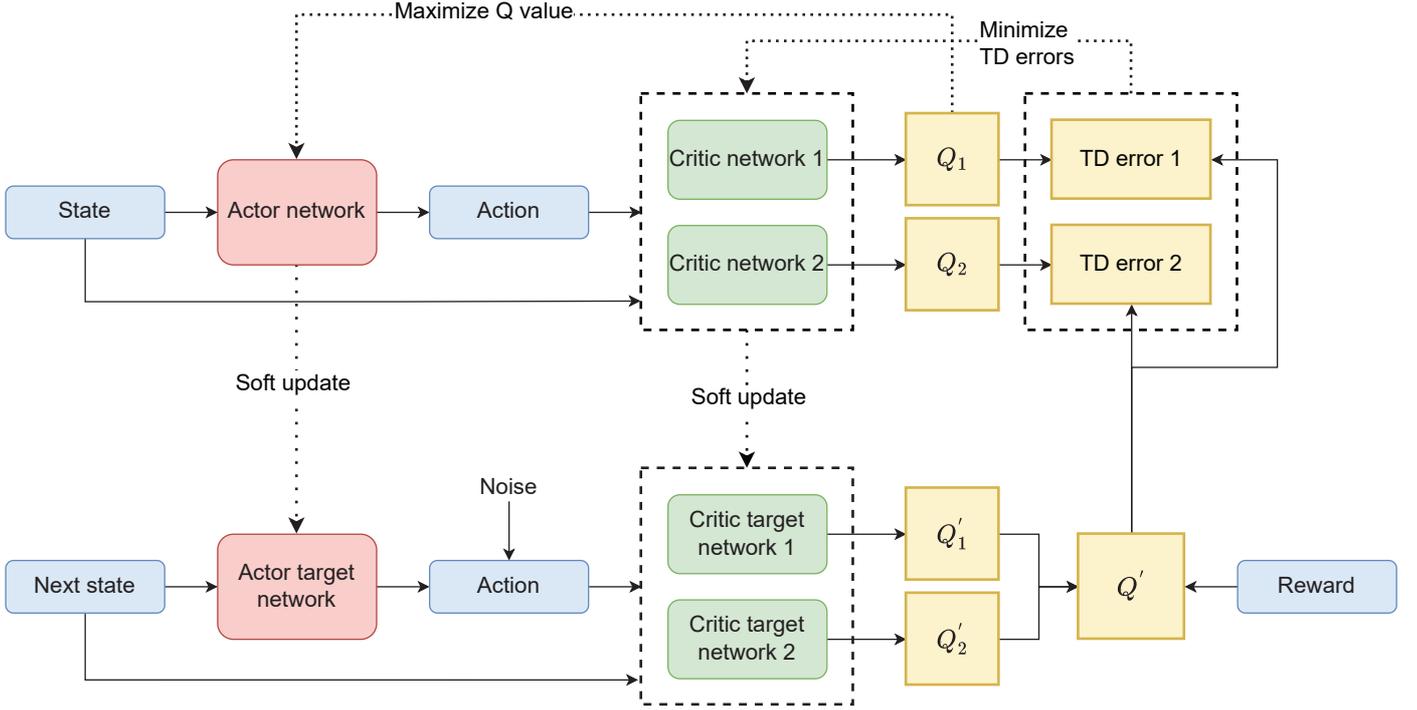}
  \caption{The process of training GAC}
  \label{fig:trainingGAC}
\end{figure*}

The process of training the proposed GAC is described in Figure \ref{fig:trainingGAC}, which is inspired by \cite{Fujimoto2018Addressing}. The GAC contains an actor network and a pair of critic networks, and each network also has a target network with the same structure. Note that the actor network adopts the structure described in Figure \ref{fig:structure}. Although critic networks adopt the same structure as the actor network, they require one additional input, i.e., action $\boldsymbol{a}_{t}$. The action will be fed into the concatenation layer directly to concatenate it with $\boldsymbol{x}^*_{in}$ and $\boldsymbol{x}^*_{out}$. Suppose we collect a batch of transitions from the replay buffer, i.e., a batch of $(S_t, \boldsymbol{a}_t,S_{t+1},R_t)$. The action $\boldsymbol{a}_t$ and the state $S_t$ would be fed into the two critic networks to calculate $Q_1$ and $Q_2$, respectively. Meanwhile, the actor target network would output an action $\boldsymbol{a}_{t+1}$ based on the next state $S_{t+1}$. The action $\boldsymbol{a}_{t+1}$ would be added with a noise $\epsilon\thicksim \mathcal{N}(0, 0.1)$, and the action with noise and $S_{t+1}$ would be fed into the critic target networks to calculate $Q'_1$ and $Q'_2$. Subsequently, the target Q value can be calculated by using Equation \ref{equ:target_Q}, where $R_t$ is the step reward at time step $t$, $\gamma$ is the discount factor, and $\min(Q'_1,Q'_2)$ would return the minimum value from $Q'_1$ and $Q'_2$.

\begin{equation}\label{equ:target_Q}
Q' = R_t + \gamma \cdot \min(Q'_1,Q'_2)
\end{equation}

The target Q value next would be used to train the pair of critic networks, where the objective is to minimize the TD error between $Q_1$, $Q_2$, and $Q'$, as described in Equation \ref{equ:train_critic}, where $n$ denotes the batch size, and $\theta_i$ represents parameters in $i$-th critic network. Using this method to update critic networks can help fend off overestimation issues, making the update smoother.

\begin{equation}\label{equ:train_critic}
\theta_i = \min_{\theta_i} \frac{\sum(Q'-Q_i)^2}{n}, i\in\{1,2\}
\end{equation}

The actor network would be trained by using a deterministic policy gradient, as described in Equation \ref{equ:train_actor}, where the objective is to maximize the Q value. Note that the actor network is updated less frequently than the pair of critic networks. This can help effectively reduce the volatility caused by the scenario in that the policy update changes the target.

\begin{equation}\label{equ:train_actor}
\phi = \max_{\phi} Q_1(S_t,\boldsymbol{a}_t)
\end{equation}

At last, the target networks would be updated by using soft update as described in Equations \ref{equ:su_1} and \ref{equ:su_2}, where $\tau$ is the soft update rate.

\begin{equation}\label{equ:su_1}
\theta'_i = \tau \theta_i + (1-\tau)\theta'_i, i\in\{1,2\}
\end{equation}
\begin{equation}\label{equ:su_2}
\phi' = \tau \phi + (1-\tau)\phi'
\end{equation}

\begin{algorithm}[!t]
  \caption{GAC Training}
  \label{alg:train}
  \KwIn {Number of episodes $EP$, number of time steps $T$, number of episodes for exploration $EXP$, update frequency $UF$, budget $B$, in-adjacency matrix $\boldsymbol{A}^{in}$, out-adjacency matrix $\boldsymbol{A}^{out}$}
  	Initialize actor network $\pi_\phi$, and critic networks $Q_{\theta_1}$ and $Q_{\theta_2}$ with random parameters $\phi$, $\theta_1$, $\theta_2$;\\
  	Initialize target networks $\phi' \leftarrow\phi$, $\theta'_1\leftarrow\theta_1$, $\theta'_2\leftarrow\theta_2$;\\
  	Initialize replay buffer $\mathcal{B}$;\\
  	\For {$episode=1\ to\ EP$}{
  		Incentivize all users with no incentive and observe users' behavior to initialize feature matrix $\boldsymbol{F}$;\\
  		$S_1=(\boldsymbol{A}^{out},\boldsymbol{A}^{in},\boldsymbol{F})$;\\
  		$\omega:=|V|^{-1}\cdot|\{v_i|v_i\in V, b_{v_i,0}==z^*\}|$;\\
  		\For {$t=1\ to\ T$}{
  			
  			\If {$episode \leq EXP$}{
  				$\boldsymbol{a}_t \thicksim \mathcal{N}(0,1)$;\\
  			}
  			\Else{
  				$\boldsymbol{a}_t \thicksim \pi_\phi(S_t) + \epsilon, \epsilon\thicksim \mathcal{N}(-\omega,1)$ ;\\
  			}
  			$\boldsymbol{a}'_t := 2^{-1} \cdot (\boldsymbol{a}_t + 1)$;\\
  			$R_t$, $L_t$ = $Assign Incentive(\boldsymbol{a}'_t,B)$ (Algorithm \ref{alg:1}) ;\\
  			$\omega:=|V|^{-1}\cdot|\{v_i|v_i\in V, b_{v_i,0}==z^*\}|$;\\
  			Observe step reward $R_t$ and create new Feature matrix $\boldsymbol{F'}$ based on $L_t$;\\
  			$S_{t+1}=(\boldsymbol{A}^{out},\boldsymbol{A}^{in},\boldsymbol{F'})$;\\
  			Store transition tuple $(\boldsymbol{S}_t,\boldsymbol{a}_t,R_t,\boldsymbol{S}_{t+1})$ in $\mathcal{B}$;\\
  			\If {$episode > EXP$}{
  				Sample a batch of transitions from $\mathcal{B}$;\\
  				Update the pair of critic networks using Equation \ref{equ:train_critic};\\
  				\If {$t$ \% $UF==0$}{
  					Update actor network using Equation \ref{equ:train_actor};\\
  					Update target networks using Equations \ref{equ:su_1} and \ref{equ:su_2};\\
  				}
  				
  			}
  		}
  	}
 
\end{algorithm} 

Algorithm \ref{alg:train} demonstrates the training process for the proposed GAC. Before starting the training process, we initialize actor and critic networks of GAC as well as the replay buffer $\mathcal{B}$. At the beginning of every episode, we assign no incentive to all users and obtain the observation of their behaviors to generate node features matrix $\boldsymbol{F}$. $\omega$ in Lines 6 and 14 represents the ratio of engaged users in the network. The state is represented by $S_t=(\boldsymbol{A}^{out},\boldsymbol{A}^{in},\boldsymbol{F})$. In the initial $EXP$ episodes, the action is captured from the normal distribution $\mathcal{N}(0,1)$ for pure exploration. After $EXP$ episodes, $\boldsymbol{a}_t$ is generated by GAC directly. To keep the approach exploring, we add noise $\epsilon$ on the generated $\boldsymbol{a}_t$, where $\epsilon\thicksim \mathcal{N}(-\omega,1)$. We capture noise from this distribution because we want to decrease the incentive allocated to users when most users in the network have been incentivized. At that moment, most users would be exerted influence by neighbors affecting them to choose $z^*$, and then they might be indirectly incentivized by their neighbors. In Line 13, the generated action $\boldsymbol{a}_t$ would be scaled to the range of [0,1]. The scaled action and budget $B$ would then be inputted to the environment introduced in Algorithm \ref{alg:1}. The environment returns the step reward $R_t$ and users' behavior log $L_t$. The received $L_t$ is used to update $\omega$ and create a new user features matrix. At every time step, we store the experience tuple $(\boldsymbol{S_t},\boldsymbol{a}_t,R_t,\boldsymbol{S_{t+1}})$ in the replay buffer $\mathcal{B}$. To save memory, we do not save the adjacency matrices to the replay buffer, as the network topology would not change in this study. After $EXP$ episodes, we sample a batch of transitions from the replay buffer to train parameters in both actor and critic networks of GAC. At every time step, the pair of critic networks would be updated by using Equation \ref{equ:train_critic}. The actor network would be updated by using a delayed update policy, i.e., it would be updated at every $UF$ time steps. After updating the actor network, the soft update will be applied to update target networks.

\section{Experimental Setup}\label{sec:expsetup}
In this section, we introduce the experimental setup for evaluating GAC. The PyTorch implementation of GAC is available on Github\footnote{https://github.com/Konatanaya/Geometric-Actor-Critic}.
\subsection{Data Preparation}
To evaluate the performance of the proposed GAC, the following three datasets are used to deploy social networks: 
\begin{itemize}
	\item \textbf{Dolphins}\footnote{http://networkrepository.com/soc-dolphins.php} dataset represents a social network of bottlenose dolphins, where a node represents a dolphin, and an edge represents frequent associations between dolphins \cite{Lusseau2003Bottlenose}. This network contains 62 nodes and 159 edges in total, and the average degree for each node is 5.1.
	
	\item \textbf{Twitter}\footnote{https://snap.stanford.edu/data/ego-Twitter.html} dataset contains 973 directed networks, 81,306 users and 1,768,149 edges in total \cite{Leskovec2012Learning}. To diminish the running time,
a sub-network that contains 236 users and 2,478 edges is selected. The average degree of the sub-network is 21.0.

	\item \textbf{Wiki-Vote}\footnote{http://networkrepository.com/soc-wiki-Vote.php} dataset contains all the voting data from the inception of Wikipedia till January 2008 \cite{Leskovec2010Signed}. This dataset contains 889 nodes and 2,914 edges, where each directed edge from node $i$to node $j$ represents that user $i$ voted user $j$. The average degree is 6.6. 
\end{itemize}

\begin{table}[!t]
\centering
\caption{Statistics of datasets}
\label{tab:datasets}
\begin{tabular}{llll}
\hline
Dataset   & $|V|$ & $|E|$ & Avg. Degree \\ \hline
Dolphins  & 62    & 159   & 5.1         \\
Twitter   & 236   & 2,478  & 21.0        \\
Wiki-Vote & 889   & 2,914  & 6.6         \\ \hline
\end{tabular}
\end{table}

To simulate user agents' behaviors in the ADM, we also assign random preferences $p_{v_i, z_m}, \forall z_m\in Z$ for all users in these three datasets. We suppose that four behavior options can be selected, and $z_0$ is regarded as the expected action $z^*$. The influence strength associated with each edge is assigned a random value from 0 to 1. Also, the sum of influence coming from a user's all in-neighbors cannot exceed 1, i.e., $\sum_{v_j\in N^{in}_{v_i}}{w_{ji}}\leq 1, \forall v_i\in V$. The statistics of three datasets are listed in Table \ref{tab:datasets}.

\subsection{Baseline Approaches}
The performance of the proposed GAC is evaluated by comparing it with the following approaches:
\begin{itemize}
\item \textbf{No Incentive} approach implies that no incentive would be allocated to users. Namely, users' decision-making would be only affected by user preferences and social influence from in-neighbors.

\item \textbf{Uniform Allocation} is a Na\"ive approach, which averagely allocates incentives based on the number of users.

\item \textbf{DGIA-IPE} focuses on allocating incentives to users in social networks. This approach consists of two components, DGIA and IPE \cite{Wu2019Adaptive}. Based on observation of users' behaviors, DGIA can estimate users' sensitivity to incentives, and IPE can estimate influential relationships among users. The estimating results from DGIA and IPE would be used to determine incentives for each user. We set up DGIA-IPE by using the same parameters used in the original paper.

\item \textbf{DBP-UCB} is a dynamic pricing algorithm to incentivize active users in the system to reposition sharing-bikes to suitable stations if necessary \cite{Singla2015Incentivizing}. Different from DGIA and the proposed GAC, DBP-UCB divides incentives into a finite set of options. The parameters of DBP-UCB would be updated based on the observation of users' behaviors. Since DBP-UCB adopts discrete price option, we give one hundred price options from 0 to 1 with an interval of 0.01.

\item \textbf{K-MAB} is an approach that can learn users' cost distribution based on users' behaviors and dynamically adjust incentive allocation policies in different scenarios \cite{Zhou2022klevel}. To better deploy K-MAB, we set one hundred price options from 0 to 1 with an interval of 0.01, which is the same as the settings of DBP-UCB. K-MAB would learn the probability that a user would accept an incentive or not. The probabilities for all users are initially set as 0.5.
\end{itemize}

The primary goal of incentive allocation is to maximize the effect of a limited budget in user incentivization. Thus, in the experiments, the performance of approaches would be primarily evaluated by this metric, i.e., the number of users who take $z^*$. This metric is default unless a clarification is given.

\subsection{System Settings}
\begin{table}[t]
\centering
\caption{Hyperparameters of GAC}
\label{tab:hyperparameters}
\resizebox{\columnwidth}{!}{
\begin{tabular}{ll}
\hline
Hyperparameters                                          & Value \\ \hline
The number of units in GraphSages and DIFFPOOLs        & 32    \\
The number of units in fully connected layers                & 64    \\
The number of clusters generated by the first DIFFPOOL  & 16    \\
The number of clusters generated by the second DIFFPOOL & 1     \\
Learning rate for actor networks                         & 3e-4  \\
Learning rate for critic networks                        & 3e-3  \\
Batch size                                               & 256   \\
Size of Replay buffer $\mathcal{B}$                                       & 1e5   \\
Update frequency $UF$                                        & 2     \\
Soft update rate $\tau$                                                   & 1e-3 \\
Discount Factor $\gamma$ & 0.99\\
Number of episodes for exploration $EXP$                                                    & 1,000      \\
Number of episodes for training GAC                     & 1e4   \\
Time steps in an episode when training GAC               & 10    \\
Time steps in an episode when evaluating GAC             & 150   \\ \hline
\end{tabular}
}
\end{table}

We conduct simulation-based experiments to evaluate the proposed GAC by comparing it with the other four baseline approaches. For GAC and its variants, we set the number of units in GraphSages and DIFFPOOLs as 32 and the number of units in fully connected layers as 64. Meanwhile, we use the first DIFFPOOL layer to assign all users to 16 clusters and the second DIFFPOOL layer to assign these 16 clusters to 1 cluster. The proposed GAC is trained by using Adam optimizer \cite{kingma2014adam}, where the learning rates for Actor and Critic networks are 3e-4 and 3e-3, respectively. We set $\tau$ used for the soft update as 1e-3 and the discount factor as 0.99. The update frequency is set as 2, i.e., the process of training the actor target network and the soft update would occur every two episodes. The size of the replay buffer is set as 1e5, and the batch size of samples is set as 256. The GAC and its variants are trained at 1e4 episodes, with 10 time steps in every episode by default. The initial 1,000 episodes are used for pure exploration, where the action is generated by using normal distribution $\mathcal{N}(0,1)$. At last, the model parameters that can generate the most optimal incentive allocation policy would be kept for evaluation. In the evaluation, the simulation for each approach lasts 150 time steps, and the budget will be refilled at the beginning of each time step. The hyperparameters of the proposed GAC are listed in Table \ref{tab:hyperparameters}.

\section{Experimental Results}\label{sec:expresult}

\begin{figure}[t]
  \centering
  \subfigure[Dolphins, $B$=3]{\includegraphics[width=0.49\columnwidth]{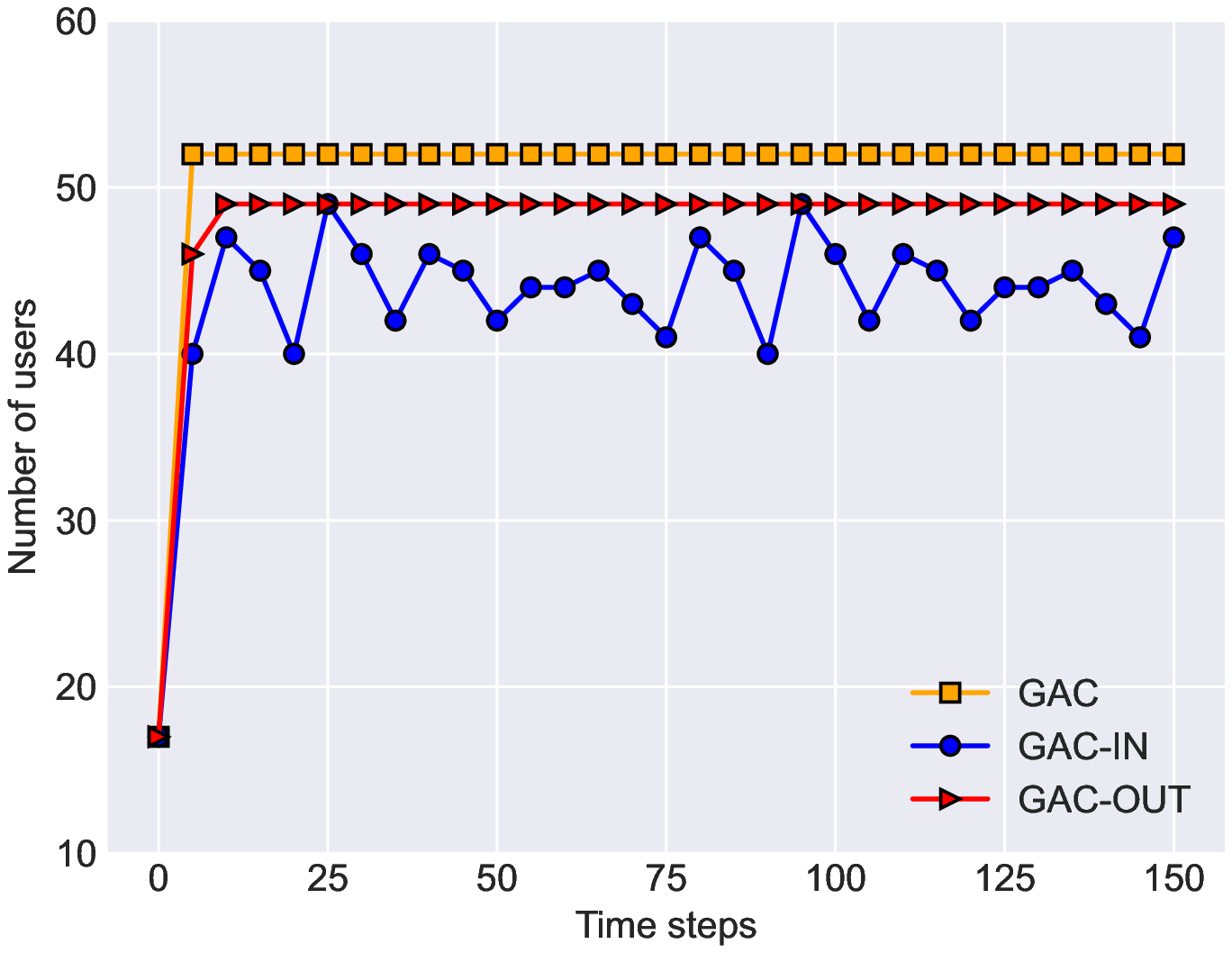}}
  \subfigure[Twitter, $B$=20]{\includegraphics[width=0.49\columnwidth]{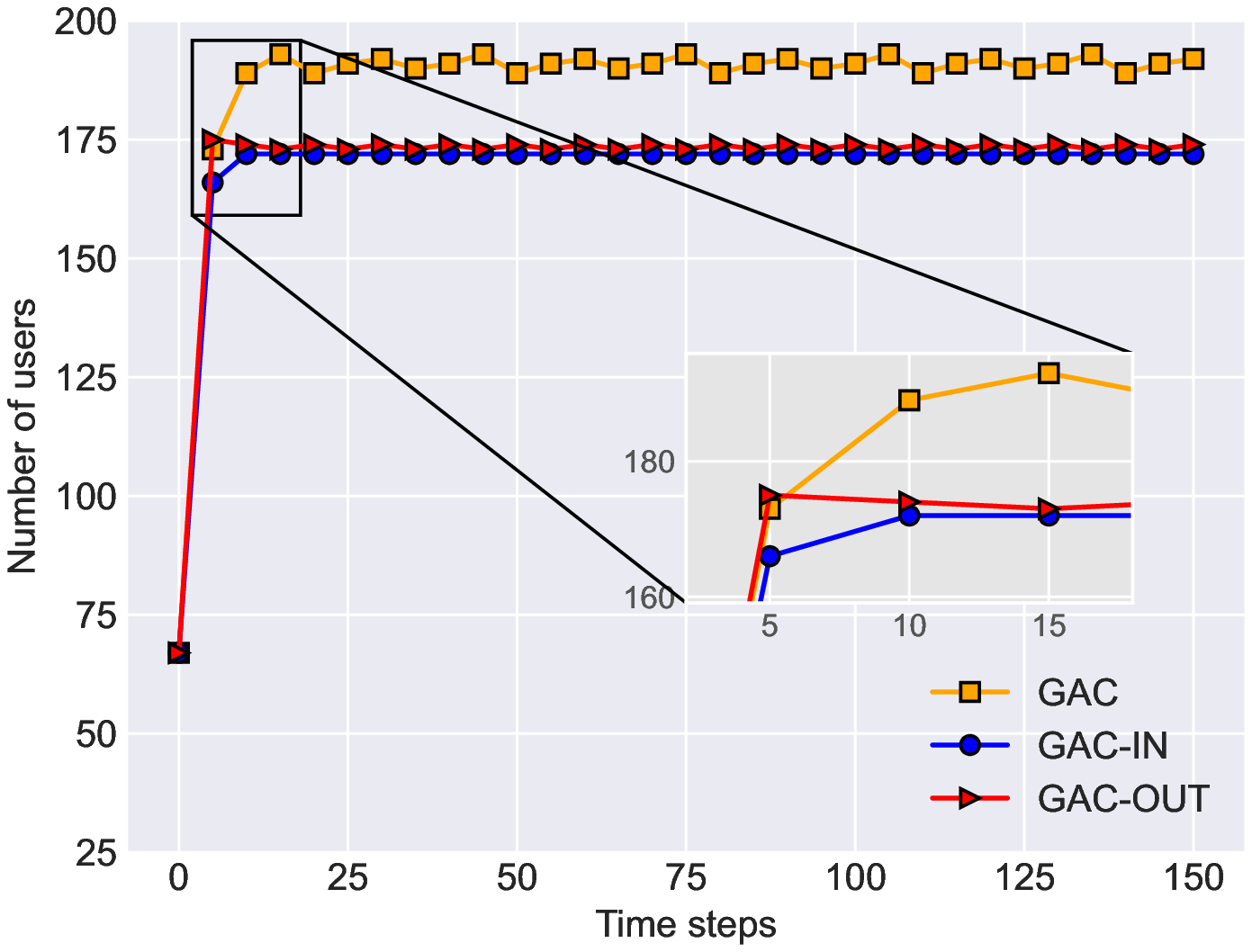}}
  \caption{Performance comparison of GAC and its variants}
  \label{fig:variants}
\end{figure}

\subsection{Impact of GAC's Structure}
In Section \ref{sec:methodology}, we explain that the purpose of inputting both in-adjacency and out-adjacency matrices is to prevent losing important information on the network. To prove the importance of this operation, we compare the GAC with its two simplified versions, i.e., GAC-IN and GAC-OUT. The difference between these two simplified GACs is the input matrices. Besides the feature matrix, GAC-IN only takes the in-adjacency matrix as input, while GAC-OUT requires the out-adjacency matrix. 

We compare the performance of these three variants by using the Dolphins and Twitter datasets, respectively. As we can observe from Figure \ref{fig:variants}, GAC sightly outperforms the simplified GAC in both datasets. Although the performance of GAC-OUT is slightly worse than GAC, it eventually converges to a stable stage. By contrast, GAC-IN performs the worst in the Dolphins dataset. A possible reason is that GAC-IN only considers users' in-neighbors when generating incentives, such that the influential users in the network are difficult to be identified. However, GAC-IN performs similarly to GAC-OUT in the Twitter dataset. Although GAC-OUT reaches the convergence faster than GAC-IN, the gap between them can be ignored after convergence. It is because the average degree in the Twitter network is higher than the Dolphins network, and users in the Twitter network who can influence others might also be affected by their neighbors.

\subsection{Impact of Noise Distribution}
\begin{figure}[t]
  \centering
  \subfigure[Dolphins, $B$=3]{\includegraphics[width=0.49\columnwidth]{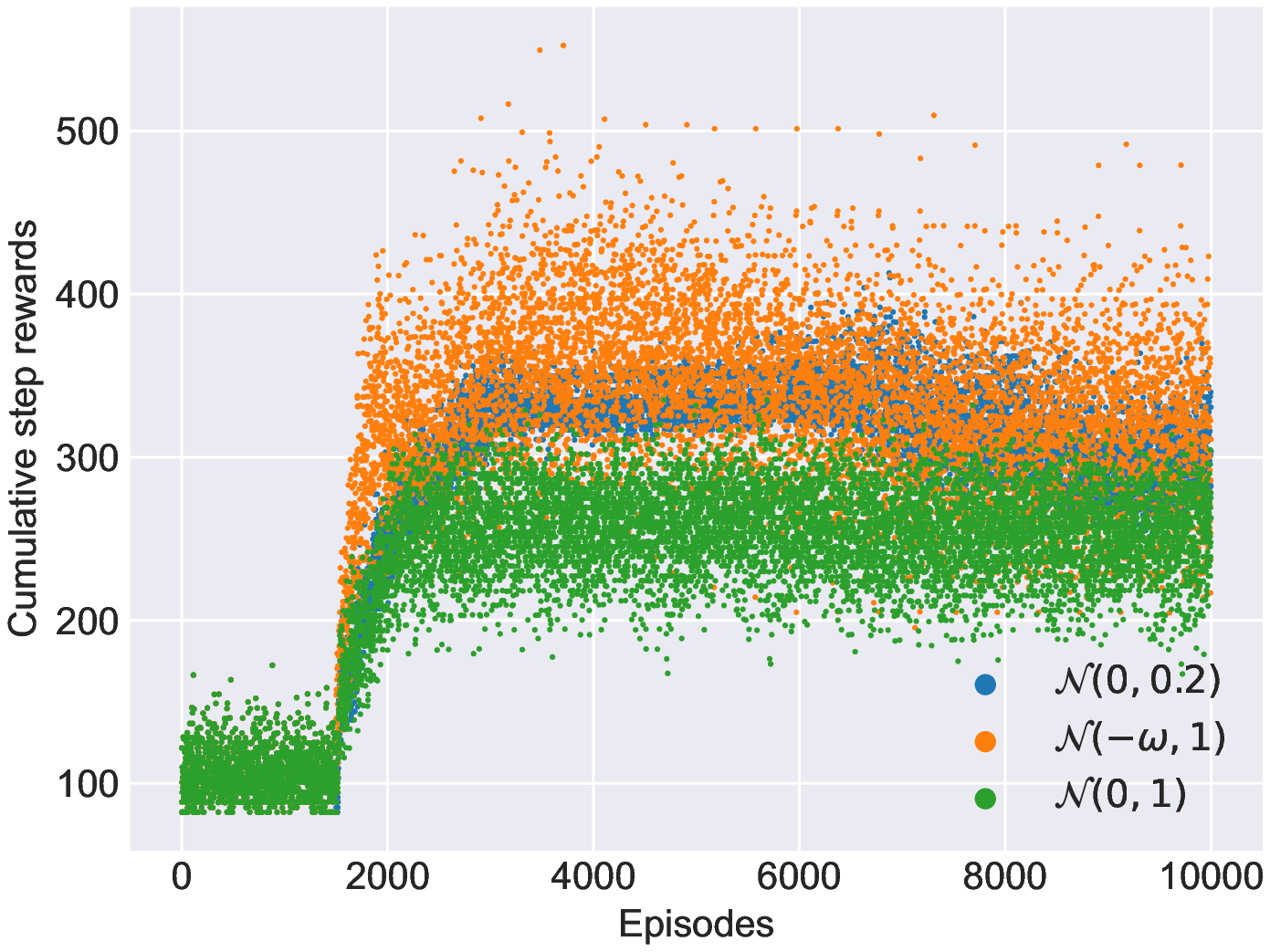}\label{fig:exploration_dolphins_r}}
  \subfigure[Twitter, $B$=20]{\includegraphics[width=0.49\columnwidth]{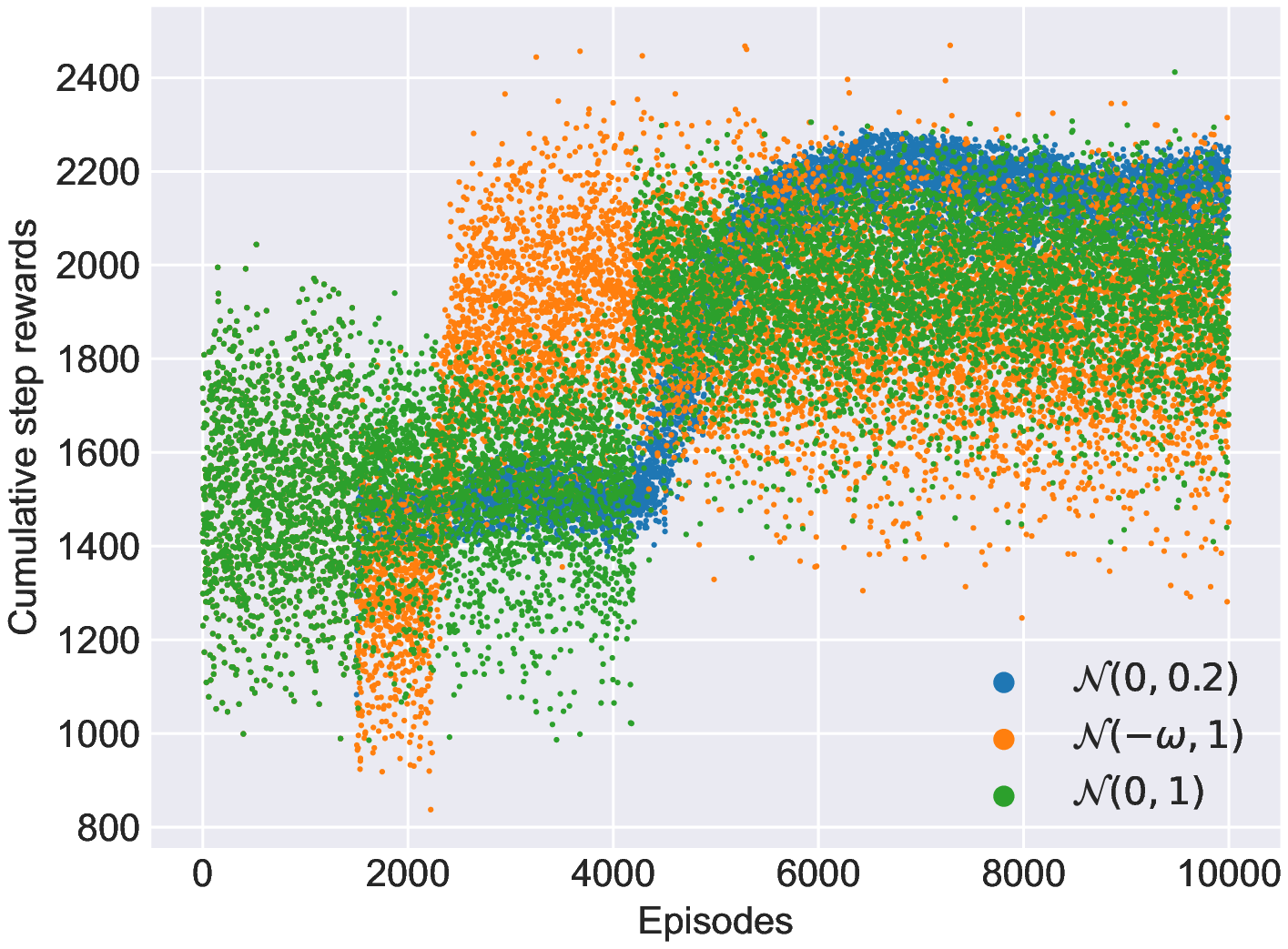}\label{fig:exploration_twitter_r}}
  \subfigure[Dolphins, $B$=3]{\includegraphics[width=0.49\columnwidth]{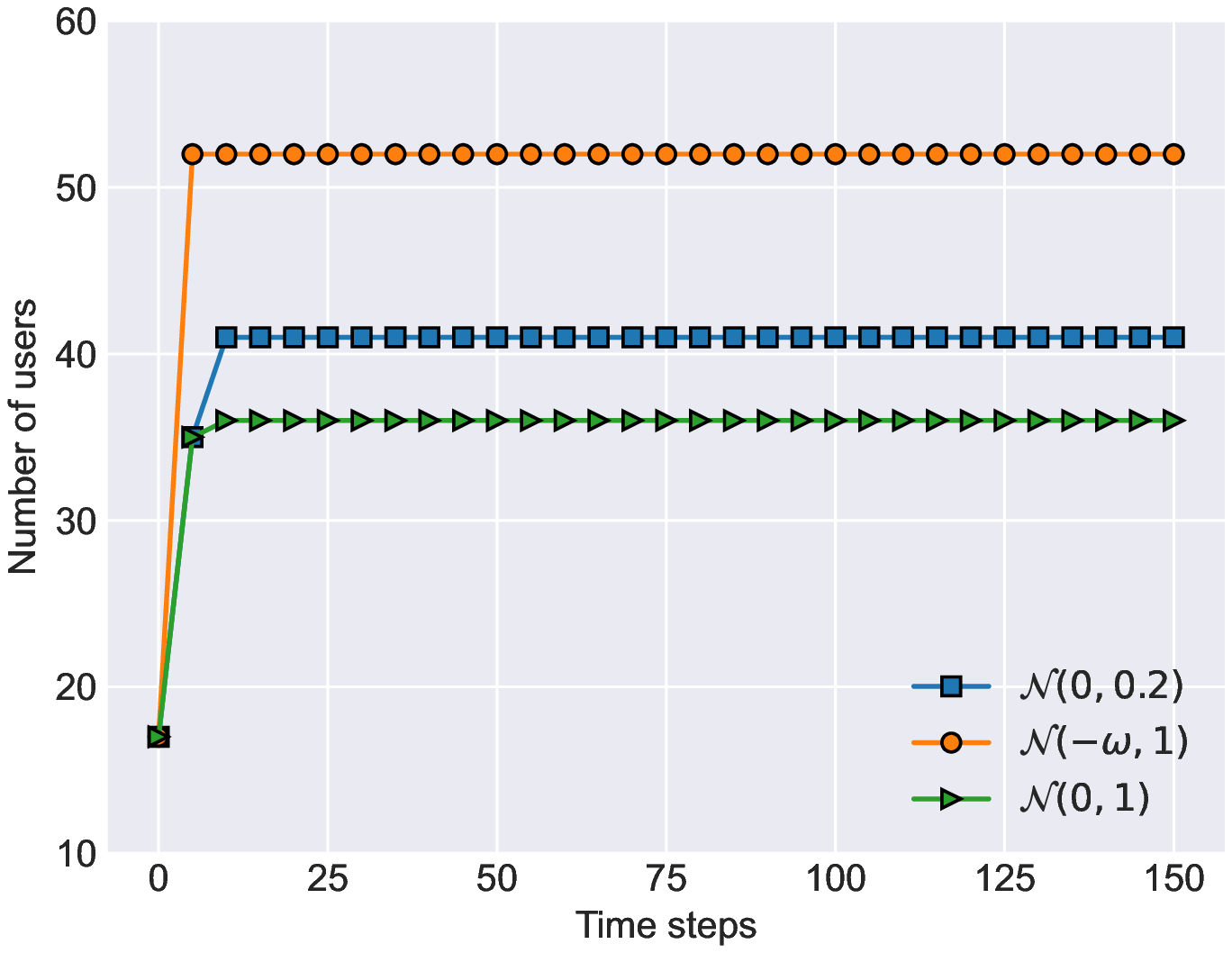}\label{fig:exploration_dolphins_num}}
  \subfigure[Twitter, $B$=20]{\includegraphics[width=0.49\columnwidth]{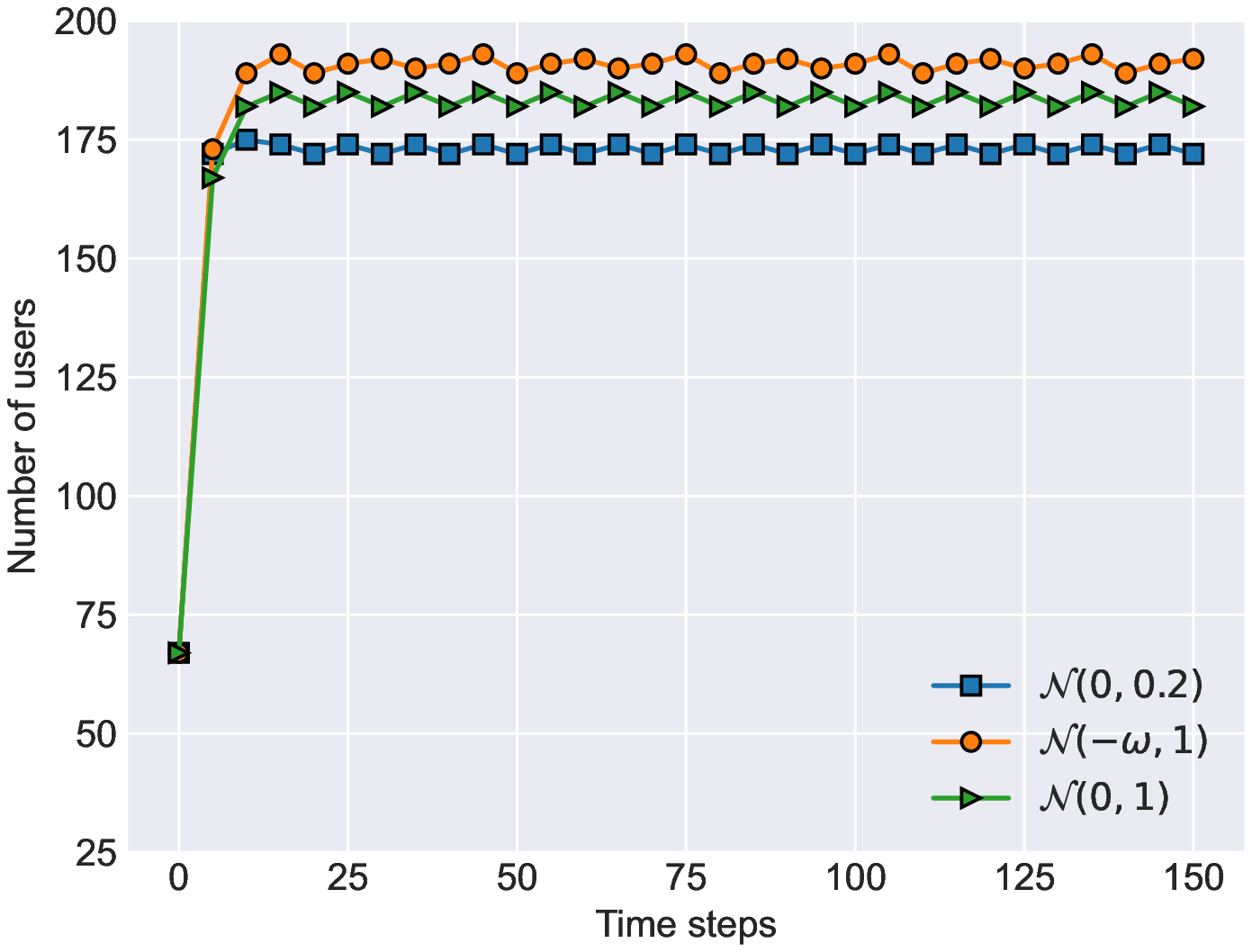}\label{fig:exploration_twitter_num}}
  \caption{Performance comparison of adding different noise}
  \label{fig:exploration}
\end{figure}

\begin{figure*}[!t]
\subfigure[Dolphins, $B$=3]{\includegraphics[width=0.33\textwidth]{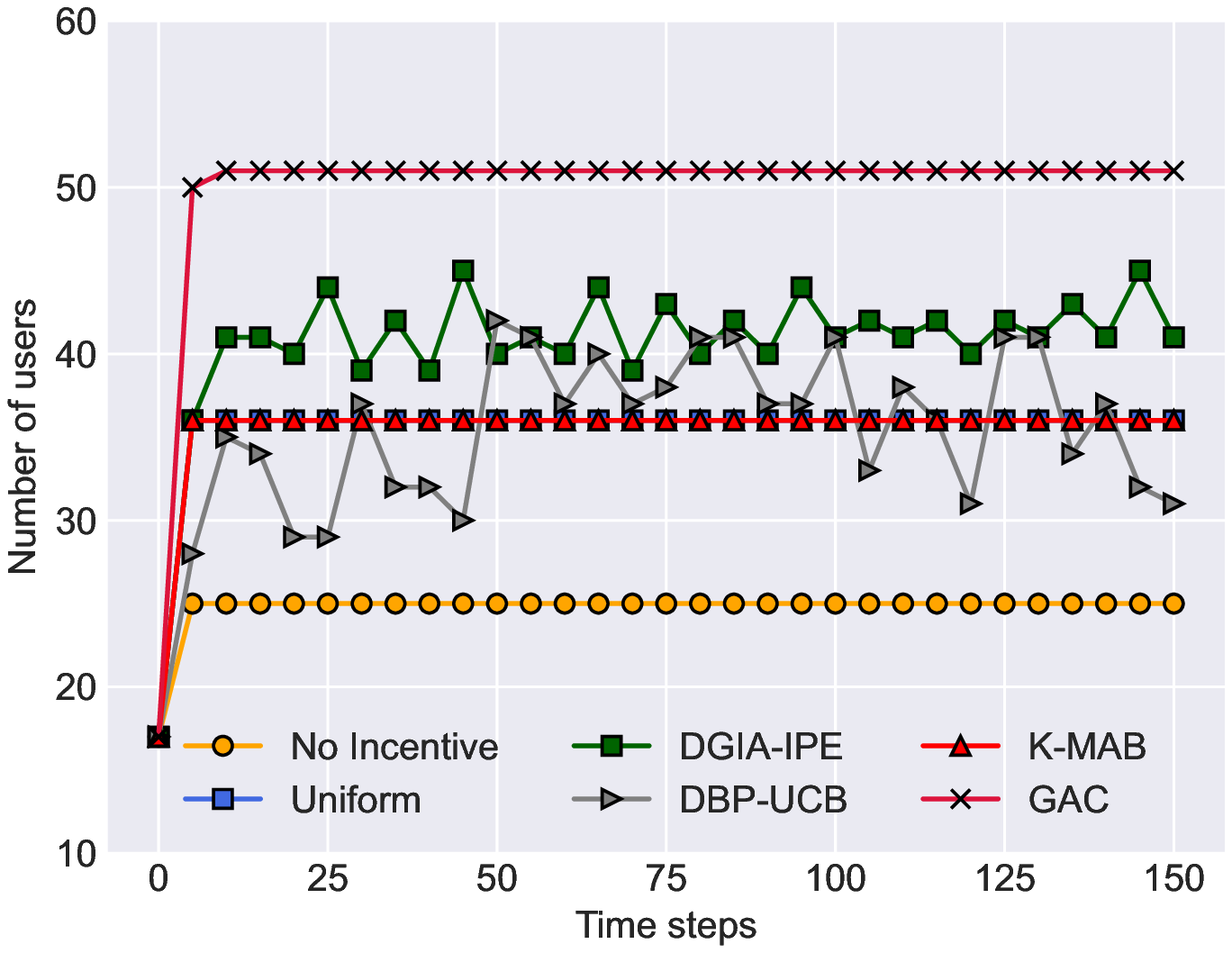}\label{fig:ex3_dolphins}}
\subfigure[Twitter, $B$=20]{\includegraphics[width=0.33\textwidth]{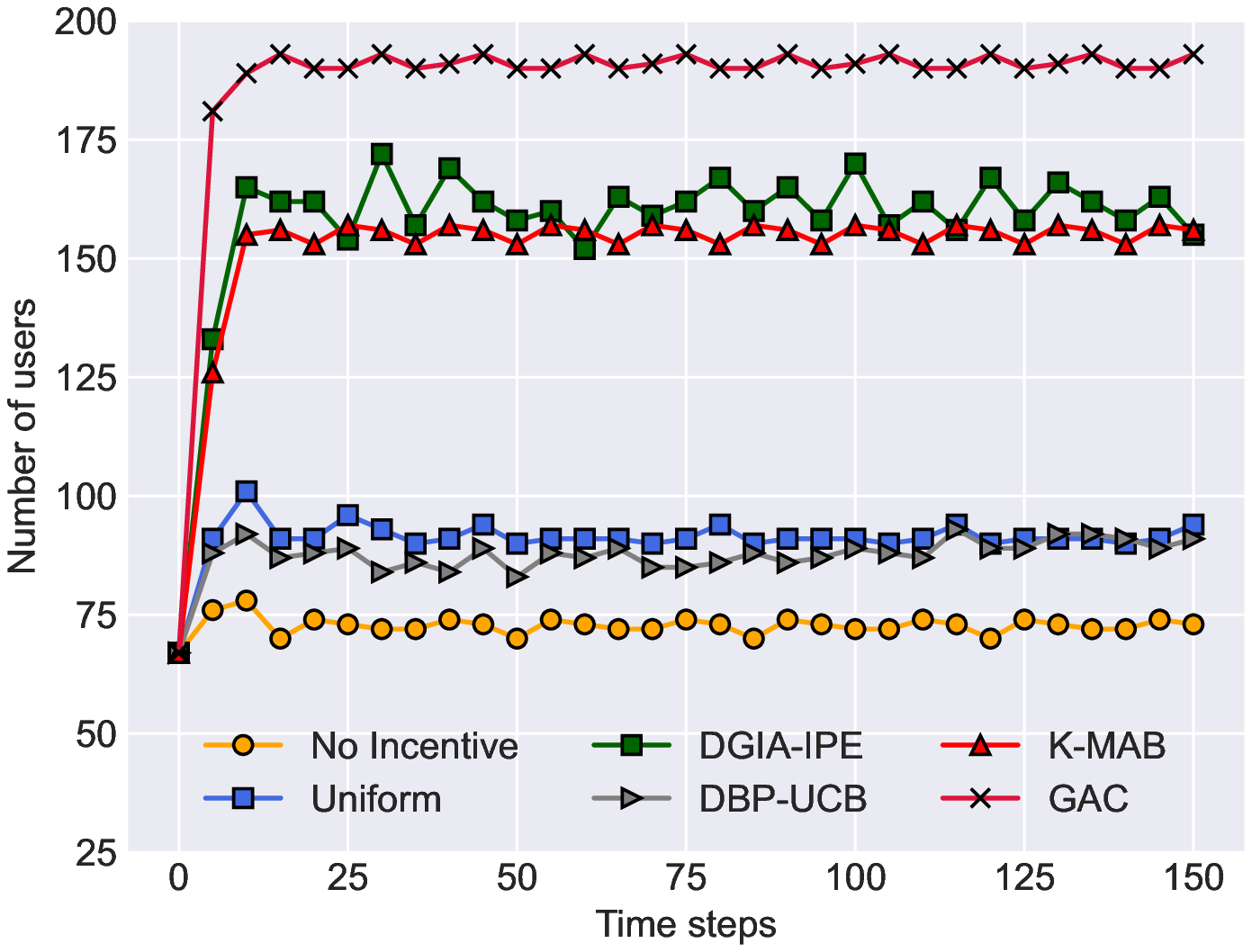}\label{fig:ex3_twitter}}
\subfigure[Wiki-Vote, $B$=40]{\includegraphics[width=0.33\textwidth]{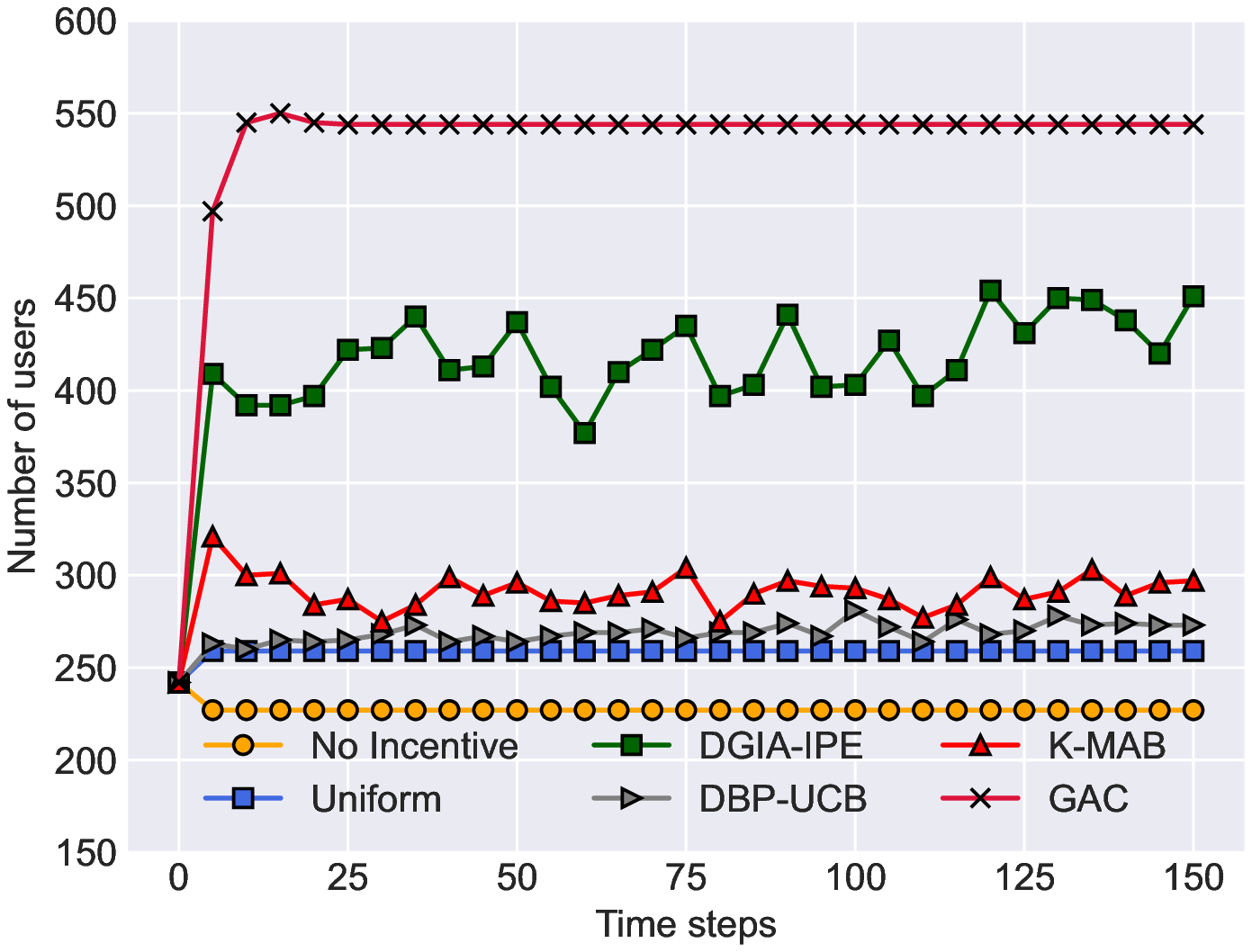}\label{fig:exp3_wiki}}
\label{fig:ex3}
\caption{Performance comparison of GAC and baseline approaches}
\end{figure*}

Many algorithms for continuous control, such as TD3 \cite{Fujimoto2018Addressing}, add noise captured from a normal distribution $\mathcal{N}(\mu,\sigma^2)$ to the policy, where $\mu$ is typically set as 0, and $\sigma$ is a fixed value. By applying this noise, the algorithm can explore different policies and eventually obtains success. However, we found that setting $\mu$ as 0 cannot help to explore effective policies in the incentive allocation problem where social influence exists. The reason is that, with the increasing number of users who choose $z^*$ in the network, more users start affecting their neighbors to select $z^*$. At that moment, it could be unnecessary to allocate incentives to some users, as they can be incentivized by the influence generated by their neighbors.

Hence, in this study, we consider capturing noise from $\mathcal{N}(-\omega,1)$, where $\omega$ represents the ratio of users who select $z^*$ in the network. We compare the performance of GAC by using three different noise distributions, i.e., $\mathcal{N}(-\omega,1)$, $\mathcal{N}(0,0.2)$, and $\mathcal{N}(0,1)$. As we can observe from Figures \ref{fig:exploration_dolphins_r} and \ref{fig:exploration_twitter_r}, using $\mathcal{N}(-\omega,1)$ can attempt diverse incentive allocation policies and receive different cumulative step rewards from the environment. These experiences are beneficial for GAC to learn if the policy is good or not. By contrast, using $\mathcal{N}(0,0.2)$ or $\mathcal{N}(0,1)$ to generate noise makes the exploration very limited, leading to the learned experience might be very similar. 

The results from Figures \ref{fig:exploration_dolphins_num} and \ref{fig:exploration_twitter_r} also demonstrate that capturing noise from $\mathcal{N}(-\omega,1)$ is more beneficial for training GAC. It allows GAC to explore various policies, enabling the model to learn from both good and bad experiences. By contrast, using $\mathcal{N}(0,0.2)$ or $\mathcal{N}(0,1)$ to generate noise makes GAC easily learn the sub-optimal policies but fail to learn the optimal policy. Notably, $\mathcal{N}(0,1)$ outperforms $\mathcal{N}(0,0.2)$ in the Twitter network, while it underperforms $\mathcal{N}(0,0.2)$ in the Dolphins network. A possible reason is that social influence plays a more crucial role in the Twitter network than the Dolphins network. Meanwhile, making $\sigma^2$ larger in the normal distribution implies that the probability of capturing a larger noise increases, such that it is possible to explore the policies that allocate users no incentive.

\subsection{Comparison with Baseline Approaches}
This experiment evaluates the proposed GAC by comparing it with the other four approaches in three different social network datasets. Figure \ref{fig:ex3_dolphins} shows the performance of five approaches in the Dolphins network. The proposed GAC can incentivize more users compared to the other five baseline approaches. We notice that DGIA-IPE outperforms K-MAB, DBP-UCB, and Uniform allocation. It is because that DGIA-IPE tends to incentivize influential users to utilize their influence, while the other three methods consider incentivizing all users directly. 

In the Twitter network, GAC and DGIA-IPE perform much better than the other baseline approaches except K-MAB, as they consider social influence when generating incentive allocation policy. In comparison with DGIA-IPE, the proposed GAC converges rapidly and performs better. Although K-MAB can reach convergence rapidly and outperform DBP-UCB, it still underperforms DGIA-IPE. DBP-UCB has a very similar performance as the Uniform allocation. It implies that DBP-UCB is ineffective in incentivizing users in such a dense social network environment.

Figure \ref{fig:exp3_wiki} shows the performance of the approaches in the Wiki-Vote network. The proposed GAC could still outperform other compared approaches. Similarly, DGIA-IPE yields better performance than the other three methods except for GAC. Also, it is notable that the trend of the number of users incentivized by DGIA-IPE is not as stable as that of the other approaches. A possible reason is that the topological structure of the Wiki-Vote network is more sparse than the Twitter network, such that the IPE algorithm cannot well estimate the influence strength associated with each edge. Meanwhile, different from its performance in the Twitter network, DBP-UCB slightly outperforms the Uniform allocation this time. It is because that DBP-UCB focuses on providing a suitable incentive to every single user. In contrast, Uniform allocation cannot provide attractive incentives for users in a large network when the budget is very limited. While DBP-UCB still underperforms K-MAB, which implies that K-MAB can be more effective than DBP-UCB in the incentive allocation task.

\subsection{Discussion}
In the experiments, we simulate a social environment and the process of incentivization in a social network, aiming to evaluate the performance of the proposed GAC in solving the incentive allocation problem. The proposed GAC is evaluated by using three different real-world datasets. We first evaluate the impact of different architectures and noise distributions on the proposed GAC. Next, we compare the performance of the GAC with existing approaches for the incentive allocation problem. Based on the experimental results, the following insights can be reviewed:

\begin{itemize}
\item The results from Experiment 1 demonstrate that inputting the complete information of the network, i.e., the in-adjacency and out-adjacency matrices, can lead to better performance. Meanwhile, the out adjacency matrix could be more important than the in adjacency matrix in the GAC.

\item Experiment 2 shows that different noise distributions can impact the performance of incentive allocation. In particular, the GAC with a dynamic noise distribution performs better than that with a static noise distribution.

\item The results of Experiment 3 demonstrate that the GAC outperforms existing baseline approaches for incentive allocation in unknown social networks. It is shown that learning representations of both users and the network can lead to a better incentive allocation result.

\end{itemize}

\section{Conclusion and Future Work}\label{sec:conclusion}
In this paper, we propose a reinforcement learning-based framework, called Geometric Actor-Critic (GAC), to solve the incentive allocation problem in unknown social networks with a limited budget, where only information about the topological structure of the network is available. At the same time, the strength of influence and the attributes of users are not available. To solve this problem, the proposed GAC learns to represent the network from both global and local perspectives and generates incentive allocation policies based on the learned information. The trained GAC is evaluated by comparing it with other baseline approaches in three real-world social network datasets. The experimental results demonstrate that the GAC outperforms other approaches in all three datasets under a budget limitation.

Although GAC lights a potential direction for the incentive allocation problem, the shortcomings of GAC cannot be ignored. Since GAC requires two adjacency matrices and a matrix of node features as input, it consumes a lot of memory space to store them temporarily. Thus, it is impossible for GAC to handle large-scale social networks. In the future, we will conduct further research on an effective reinforcement learning-based framework for the incentive allocation problem in large-scale networks.

\bibliographystyle{elsarticle-num}
\bibliography{reference} 

\end{document}